\documentclass[11pt]{article}
\usepackage{epsfig}
\usepackage{cite}
\usepackage{amsmath,amssymb,amsfonts}
\usepackage{graphicx}
\usepackage{slashed}
\usepackage{fancyhdr}
\usepackage{braket}
\usepackage{subfigure}

\newcommand{\beq}{\begin{equation}}
\newcommand{\eeq}{\end{equation}}
\newcommand{\beqa}{\begin{eqnarray}}
\newcommand{\eeqa}{\end{eqnarray}}

\providecommand{\eins}{\mbox{{\small 1}\hspace{-0.37em}1}}

\begin{document}


\vspace{.6in}

\begin{center}

\bigskip


{{\Large\bf Threshold neutral pion photoproduction off the tri-nucleon
to ${\cal O}(q^4)$}}

\end{center}

\vspace{.3in}

\begin{center}
{\large 
Mark~Lenkewitz$^\star$\footnote{email: lenkewitz@hiskp.uni-bonn.de},
Evgeny~Epelbaum$^\ast$\footnote{email: evgeny.epelbaum@rub.de},
H.-W.~Hammer$^\star$\footnote{email: hammer@hiskp.uni-bonn.de}
 {\footnotesize and}
Ulf-G. Mei{\ss}ner$^\star$$^\ddagger$\footnote{email: meissner@hiskp.uni-bonn.de}
}

\vspace{1cm}

$^\star${\it Universit\"at Bonn,
Helmholtz--Institut f\"ur Strahlen-- und Kernphysik (Theorie)\\
and Bethe Center for Theoretical Physics,
D-53115 Bonn, Germany}

\bigskip

$^\ast${\it Ruhr-Universit\"at Bochum, Institut f\"ur Theoretische Physik
            II\\D-44870 Bochum, Germany}

\bigskip

$^\ddagger${\it Forschungszentrum J\"ulich, Institut f\"ur Kernphysik 
(IKP-3),\\ Institute for Advanced Simulation (IAS-4) and
 J\"ulich Center for Hadron Physics\\ D-52425 J\"ulich, Germany}

\bigskip

\bigskip

\end{center}

\vspace{.4in}

\thispagestyle{empty} 

\begin{abstract}\noindent 
We calculate electromagnetic neutral pion production off tri-nucleon bound
states ($^3$H, $^3$He) at threshold in chiral nuclear effective field theory
to fourth order in the standard heavy baryon counting. We show that
the fourth order two-nucleon corrections to the S-wave multipoles at threshold
are very small. This implies that a precise measurement of the S-wave cross
section for neutral pion production off $^3$He allows for a stringent test
of the chiral perturbation theory prediction for the S-wave  electric
multipole $E_{0+}^{\pi^0 n}$.
\end{abstract}

\vfill

\pagebreak

\section{Introduction}

In the absence of free neutron targets, light nuclei like the deuteron 
or three-nucleon bound states like $^3$H (triton) or $^3$He can be used
to unravel the properties of neutrons. For a recent review on extracting 
the neutron structure from electron or photon scattering off light nuclei, 
see Ref.~\cite{Phillips:2009af}. Of particular interest in this respect
is threshold neutral pion photo- and electroproduction off the nucleon.
This  is one of the finest reactions to test the chiral dynamics of QCD, see 
Ref.~\cite{Bernard:2007zu} for a recent review.  Arguably most striking is
the  counterintuitive chiral perturbation theory prediction (CHPT) that 
the elementary neutron S-wave multipole $E_{0+}^{\pi^0 n}$ is larger in 
magnitude than the corresponding one of the proton,  
 $E_{0+}^{\pi^0 p}$ \cite{Bernard:1994gm,Bernard:2001gz}. This prediction
was already successfully tested in neutral pion photo- 
\cite{Beane:1995cb,Beane:1997iv} 
and electroproduction off the deuteron \cite{Krebs:2004ir}. 
However, given the scarcity and precision of the corresponding data, 
it is mandatory to study also pion production off tri-nucleon bound states, 
that can be calculated nowadays to high precision based on chiral nuclear 
effective field theory (EFT). This framework extends CHPT to  nuclear physics 
(for  recent reviews, see \cite{Epelbaum:2008ga,Machleidt:2011zz}). 
$^3$He appears to be a particularly promising target
to extract information about the neutron amplitude. This idea is usually 
invoked for spin-dependent quantities since the $^3$He wave function is
strongly dominated by the principal S-state
component which suggests that its spin is largely driven by the one of the 
neutron.  
The photo- and electroproduction of neutral pions from tri-nucleon
systems ($^3$He and $^3$H) was considered in Ref.~\cite{Lenkewitz:2011jd} 
based on chiral 3N wave functions at next-to-leading order in the 
standard heavy baryon expansion.
Here, we extend this calculation to fourth order in the chiral
expansion, including consistently {\em all} next-to-next-to-leading order
contributions in the standard heavy baryon expansion.
This amounts to a complete 
(i.e.~subleading) one-loop calculation in the one-nucleon sector.
An investigation of the role of nucleon recoil, accounting for the
scale $\chi=\sqrt{M_\pi m_N}\simeq 340$ MeV, is beyond the scope of this
work and will be left for the future.

 Experimentally, neutral pion photoproduction off light nuclei has so 
far only been studied at Saclay~\cite{Argan:1980zz,Argan:1987dm}  
and at Saskatoon  \cite{Bergstrom:1998zz,Barnett:2008zz}. Clearly, new
measurements using CW beams, modern targets and detectors are urgently
called for. The results presented below show that we are able to calculate
neutral pion photoproduction off tri-nucleon systems to very good 
precision and, moreover,  that 
the S-wave cross section for neutral pion production off $^3$He 
is very sensitive to the elementary  $E_{0+}^{\pi^0 n}$ multipole (as already
stressed in  Ref.~\cite{Lenkewitz:2011jd}).

Our manuscript is organized as follows. Section~2 contains all the necessary
formalism. In particular, we spell out in detail the fourth order 
two-nucleon corrections that modify the third order results of
Ref.~\cite{Lenkewitz:2011jd}. We also briefly recall the contributions
worked out in that paper. Section~3 contains our results and the discussion
of these. We end with a short summary and outlook in Sec.~4.

\section{Formalism}

\subsection{Generalities}

\begin{figure}[ht]
        \begin{center}
        \includegraphics[width=0.7\linewidth]{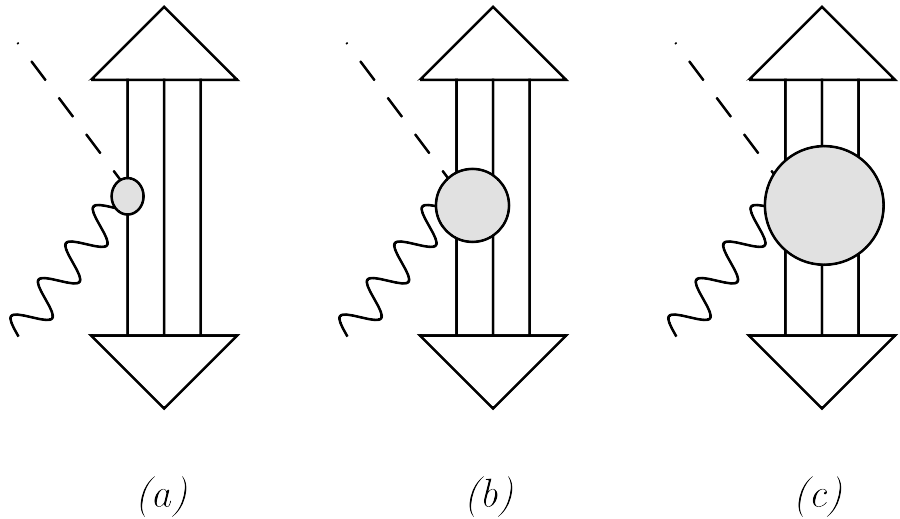}
        \end{center}
        \vspace{-0.2cm}
        \caption{Different topologies contributing to pion
        production off the three-nucleon bound state (triangle).
        (a), (b) and (c) represent the single-, two- and three-nucleon
        contributions, respectively. Solid, dashed and wiggly lines
        denote nucleons, pions and photons, in order. Topology (c) 
        does not contribute to the order considered here (NNLO).}
        \label{fig:123N}
        \end{figure}

Pion production off a tri-nucleon bound state
is given in terms of three different topologies of Feynman diagrams,
see Fig.~\ref{fig:123N}. While the single-nucleon contribution 
(a) corresponding to the standard impulse approximation
features the elementary neutron and proton production amplitudes, the
nuclear corrections are given by two-body (b) and three-body 
(c) terms.  Based on the power counting developed in \cite{Beane:1995cb},
at next-to-next-to-leading order (NNLO), only the to\-po\-lo\-gies (a) and (b)
contribute. Topology (c) starts to contribute at fourth order to 
P-wave multipoles and is thus not of relevance for our
considerations.  Here, we will specifically consider threshold photo- and
electroproduction parameterized in terms of the electric $E_{0+}$
and longitudinal $L_{0+}$ S-wave multipoles. In particular, we
study the sensitivity of the  $^3$H/$^3$He S-wave multipoles
to the elementary  $E_{0+}^{\pi^0 n}$ multipole, as the production
amplitude off the proton is well understood experimentally
and theoretically.

To analyze pion photo- and electroproduction of the tri-nucleon system, 
one has to  calculate  the nuclear matrix  element of the 
transition operator $\hat{O}$ as:
\beq\label{eq:OME}
\langle M_J^\prime|\hat{O}|M_J\rangle_{\psi}
:=\langle \psi\, M_J^\prime\, \vec{P}^\prime_{3N}\,\vec{q}\,|\,\hat{O}\,|\,\psi \,
M_J \,\vec{P}_{3N}\,\vec{k}\, \rangle\, ,
\eeq
where $\psi$ refers to the three-nucleon state
and
$\vec{k}$, $\vec{q}$, $\vec{P}_{3N}$ and $\vec{P}^\prime_{3N}$
denote the momentum of the  real or virtual
photon, produced pion and the initial
and final momenta of the  3N nucleus, respectively. 
We use the Coulomb gauge throughout this work. Note, however,
that the Coulomb gauge condition is only satisfied up to higher
order corrections in our calculation. Possible improvements are
discussed in Refs.~\cite{Djukanovic:2004px,Kvinikhidze:2009be}.

The 3N bound state has total nuclear angular momentum $J=1/2$ with magnetic
quantum numbers $M_J$ for the initial and $M_J^\prime$ for the final nuclear 
state. $J$ can be decomposed in total spin $S=1/2,3/2$ and total orbital
angular momentum $L=0,1,2$. The total isospin is a mixture of two components, 
$T=1/2$ and $3/2$. While the $T=1/2$ component is large, the small $T=3/2$
component emerges due to isospin breaking and is neglected in  
our calculation. The isospin magnetic quantum numbers
are $M_T=M_{T^\prime}=1/2$ for $^3$He and $M_T=M_{T^\prime}=-1/2$ for $^3$H.
In this paper, we consider neutral pion production by real or virtual photons
off a spin-1/2 particle - either the nucleon or the $^3$H and $^3$He nuclei.
At threshold, the corresponding transition matrix takes the form
\beq
  \label{eq:multidef}
  \mathcal{M}_\lambda =
  2i\, E_{0+} \, (\vec{\epsilon}_{\lambda, \text{T}}\cdot\vec{S} )
  + 2i \, L_{0+}\, (\vec{\epsilon}_{\lambda, \text{L}}\cdot\vec{S}) \,,
\eeq
with $\vec{\epsilon}_{\lambda, \text{T}} = \vec{\epsilon}_\lambda
-(\vec{\epsilon}_\lambda\cdot\hat{k})\hat{k}$ and $\vec{\epsilon}_{\lambda,
  \text{L}} = (\vec{\epsilon}_\lambda\cdot\hat{k})\hat{k}$ the transverse and
longitudinal photon polarization vectors, and $\vec{S}$ is the spin vector. 
The transverse and 
longitudinal  S-wave multipoles are denoted by $ E_{0+}$ and $L_{0+}$,
respective\-ly. Note that  $L_{0+}$  contributes only for virtual photons.
Also, as will be explained later in more detail, due to boost effects
that appear at next-to-next-to-leading order (NNLO), there will be P-wave
corrections to the S-wave multipoles. In the following, we will first
recapitulate the leading one-loop calculation from
Ref.~~\cite{Lenkewitz:2011jd} and then
systematically work out the NNLO corrections.

The chiral expansion of the pertinent amplitudes is done in complete analogy
to the case of the deuteron discussed in detail in Ref.~\cite{Beane:1997iv}.
We summarize  here only briefly the most important issues. First, the
production operator on the single nucleon is worked out to fourth order
in the chiral expansion. This is consistent with the two-body contributions
(exchange currents) worked out here. The leading two-nucleon terms start at third order
and have been evaluated in Ref.~\cite{Lenkewitz:2011jd}. At fourth order,
there are further corrections to the two-body currents which can be classified as boost,
static and recoil contributions. These will be worked out in detail in the
following paragraphs. Second, using the same arguments as in
Ref.~\cite{Beane:1995cb,Beane:1997iv}, 
one can show that there are no contributions from
 short-distance  four-nucleon-pion-photon operators at threshold
 (which is the case we consider here).  Notice that such operators do
 contribute at fourth order for nonvanishing momenta of the produced pion. 

\subsection{Single nucleon and leading two-nucleon contributions at 
threshold}

As explained before, the matrix element Eq.~(\ref{eq:OME}) receives
contributions from one- and two-nucleon operators at the order we are working. 
Consider first the single 
nucleon contribution, given in terms of the 1-body transition operator
$\hat{O}^{\rm 1N}$. After some algebra, one finds
\beq\label{eq:O1fin}
\langle M_J^\prime|\hat{O}^{\rm 1N}|M_J\rangle_{\psi}  =
i \vec{\epsilon}_{\lambda,\text{T}}\cdot\vec{S}_{M_J^\prime
  M_J}\Big(  E_{0+}^{\pi^0p} F_T^{S+V} + E_{0+}^{\pi^0n}  F_T^{S-V}\Big)
+ i \vec{\epsilon}_{\lambda,\text{L}}\cdot\vec{S}_{M_J^\prime
  M_J}\Big( L_{0+}^{\pi^0p}
F_L^{S+V}+ L_{0+}^{\pi^0n} F_L^{S-V}\Big)~,
\eeq
where $F_{T/L}^{S\pm V}\equiv F_{T/L}^S\pm F_{T/L}^V$ and 
$F_{T/L}^{S,V}$ denote the corresponding form factors of the
3N bound state,
\begin{align}
\label{eq:F1N}
F_{T/L}^S \, \vec{\epsilon}_{\lambda,\text{T/L}}\cdot\vec{S}_{M_J^\prime
  M_J}
&= \frac{3}{2} \,  \langle M_J^\prime
| \vec{\epsilon}_{\lambda,\text{T/L}}\cdot\vec{\sigma}_{1}|
M_J\rangle_{\psi}~, \nonumber \\
F_{T/L}^V  \, \vec{\epsilon}_{\lambda,\text{T/L}}\cdot\vec{S}_{M_J^\prime
  M_J}
&= \frac{3}{2} \, \langle M_J^\prime
|\vec{\epsilon}_{\lambda,\text{T/L}}\cdot\vec{\sigma}_{1}\tau_{1}^z|
M_J\rangle_{\psi}~, 
\end{align}
which parameterize the 
response of the
composite system to the excitation by photons in spin-isospin space.
$\vec{S}_{M_J^\prime M_J}$ are the corresponding spin transitions matrix elements.
In the above equation, $\vec \sigma_{i}$ ($\vec \tau_i$) denote the spin
(isospin) Pauli matrices corresponding to the nucleon $i$. Furthermore, $z$
refers to the isospin quantization axis. 

Using the 3N wave functions from chiral nuclear EFT at the appropriate order,
the pertinent matrix elements in Eq.~(\ref{eq:F1N}) can be evaluated. 
Here, we use  chiral 3N wave functions obtained from the N$^2$LO
interaction in the Weinberg power counting
\cite{Epelbaum:2003gr,Epelbaum:2003xx}.\footnote{The consistency of the
  Weinberg counting for short-range operators and
  the non-perturbative renormalization of chiral EFT 
  are currently under discussion, see the 
  reviews~\cite{Epelbaum:2008ga,Machleidt:2011zz} and references therein 
  for more details.
  A real alternative to the Weinberg approach for practical
  calculations in systems of three and more nucleons, however, is not 
  available.} 
In order to estimate the 
error from higher order corrections, we use wave functions
for five different combinations of the 
cutoff $\tilde{\Lambda}$ in the spectral function  representation
of the two-pion exchan\-ge and the cutoff $\Lambda$ used to regularize the 
Lipp\-mann-Schwinger equation for the two-body T-ma\-trix.
The wave functions are taken from Ref.~\cite{Liebig:2010ki,Noggaprivate}
and the corresponding cutoff combinations in units of MeV are 
$(\tilde{\Lambda},\Lambda)$ = (450,500), (600,500),
(550,600), (450,700), (600,700).
All five sets describe the binding energies of the 
$^3$He and $^3$H nuclei equally well 
(after inclusion of the
corresponding three-nucleon force). 

The one-body contribution to the 3N multipoles are given by
\begin{align}\label{eq:mult1N}
E_{0+}^{\rm 1N}
& = \frac{K_{\rm 1N}}{2} \left( E_{0+}^{\pi^0 p} \, F_T^{S+V} + E_{0+}^{\pi^0 n}
\, F_T^{S-V}\right)~,\nonumber\\
L_{0+}^{\rm 1N}
& = \frac{K_{\rm 1N}}{2} \left( L_{0+}^{\pi^0p} \, F_L^{S+V} + L_{0+}^{\pi^0n}
\, F_L^{S-V} \right)~.
\end{align}
Here, $K_{1N}$ is the kinematical factor to account for the change in phase
space from the 1N to the 3N system, 
\begin{equation}\label{eq:kine1N}
K_{\rm 1N}
 = \frac{m_N+M_\pi}{m_{\rm 3N}+M_\pi}\frac{m_{\rm 3N}}{m_N}
\approx 1.092\,,
\end{equation}
with $m_N$ being the nucleon mass and 
$m_{\rm 3N}$ the mass of the three-nucleon bound state.
Throughout, we denote  the neutral pion mass by $M_\pi$, but the
S-wave multipoles are usually given in terms of the charged
pion mass $M_{\pi^+}$ and also the charged pion mass appears
in some of the two-nucleon contributions.

We evaluate the matrix elements for the one-body contribution
in Eq.~(\ref{eq:F1N}) numerically
with Monte Carlo integration using the VEGAS algorithm~\cite{Vegas}.
The results for the form factors $F_{T/L}^{S\pm V}$ are given
in Table~\ref{tab:FTL}.
\renewcommand{\arraystretch}{1.3}
\begin{table}[t]
\begin{center}
\begin{tabular}{|c||c|c|}
\hline 
nucleus & $^3$He & $^3$H \\
\hline
$F_{T}^{S+V}$ &  $0.017(13)(3)$ & $1.493(25)(3)$ \\ 
$F_{T}^{S-V}$ &  $1.480(26)(3)$ & $0.012(13)(3)$ \\
$F_{L}^{S+V}$ &  $-0.079(14)(8)$ & $1.487(27)(8)$ \\
$F_{L}^{S-V}$ &  $1.479(26)(8)$ & $-0.083(14)(8)$ \\
\hline
\end{tabular}
 \caption{Numerical results for the form factors $F_{T/L}^{S\pm V}$.
The first error is 
our estimation of the theoretical uncertainty resulting from the
truncation of the chiral expansion while the second one
   is the statistical error from the Monte Carlo integration.}
  \label{tab:FTL}
\end{center}
\end{table}
The first error represents the theoretical uncertainty estimated from the 
cutoff variation
in the wave functions. We take the central value defined by the five
different cutoff sets as our prediction and estimate the theory error
from higher-order corrections
from the spread of the calculated values. Strictly speaking, this procedure
gives a lower bound on the error, but in practice it generates a 
reasonable estimate.   The second error is the statistical error from 
the Monte Carlo evaluation of the integrals. It is typically much 
smaller than the estimated theory error and can be neglected.
Note that the single nucleon results in  Table~\ref{tab:FTL} are consistent
with isospin symmetry within the numerical accuracy. This 
feature was not  explicitly enforced and  provides
a check on our calculation.

We stress that we follow the nuclear EFT formulation of Lepage, 
in which the whole effective potential  is iterated to all orders when 
solving the Schr\"odinger equation for the nuclear states.
As discussed in Ref.~\cite{Lepage:1997cs}, the cutoff should be kept
of the order of the breakdown scale or below
in order to avoid unnatural scaling 
of the coefficients of higher order terms. Indeed, using larger cutoffs
can lead to a violation of certain low-energy theorems as demonstrated in 
Ref.~\cite{Epelbaum:2009sd} for an exactly solvable model.

The error related to the expansion of the
production operator is difficult to estimate given that the convergence in the
expansion for the single nucleon S-wave multipoles is known to be slow, see 
Ref.~\cite{Bernard:1994gm} for an extended discussion.
We therefore give only a rough estimate of this uncertainty. The
extractions of the proton S-wave photoproduction amplitude based
on CHPT using various approximations \cite{FernandezRamirez:2009jb} 
lead to an uncertainty $\Delta E_{0+}^{\pi^0 p} \approx \pm 0.05\times
10^{-3}/M_{\pi^+}$, which is about 5\%. The uncertainty  of the neutron 
S-wave threshold amplitude is estimated to be the same.
Consequently, our estimate of the error on the single nucleon amplitude
is 5\%.

\begin{figure}[t]
        \begin{center}
        \includegraphics[width=0.7\linewidth]{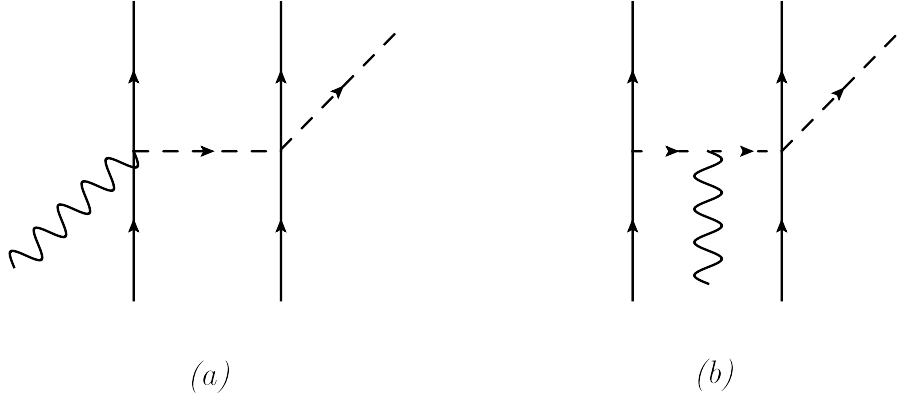}
        \end{center}
        \vspace{-0.2cm}
        \caption{Leading two-nucleon contributions to the 
         nuclear pion production matrix element at threshold.
        Solid, dashed and wiggly lines
        denote nucleons, pions and photons, in order.}
        \label{fig:abterm}
        \end{figure}

We now switch to the two-nucleon contribution. 
In Coulomb gauge, only the two Feynman diagrams shown in 
Fig.~\ref{fig:abterm} contribute at threshold 
at third order~\cite{Beane:1995cb,Beane:1997iv}.
%
%
Their contribution to the multipoles can be written as
\beqa\label{eq:mult2N}
E_{0+}^{\rm 2N} &=& K_{\rm 2N} \, (F_T^{(a)} -F_T^{(b)})~,\nonumber\\
L_{0+}^{\rm 2N} &=& K_{\rm 2N} \, (F_L^{(a)} -F_L^{(b)})~,
\eeqa
with the prefactor
\beq
K_{\rm 2N} =  \frac{M_\pi e g_A m_{\rm 3N}}{16\pi(m_{\rm 3N}+M_\pi)
(2\pi)^3 F_\pi^3}\approx 0.135\mbox{ fm }\times 10^{-3}/M_{\pi^+}~.
\eeq
The numerical value for $K_{\rm 2N}$ 
was obtained using $g_A=1.26$ for the axial-vector
coupling constant, $F_\pi=93$~MeV for the pion decay constant, and the 
neutral pion mass $M_\pi=135$~MeV. Note that we use the older values
for $g_A$ and $F_ \pi$ to be consistent with the numbers used in the evaluation
of the single nucleon multipoles \cite{Bernard:1994gm,Bernard:2001gz}.
The transverse and longitudinal form factors $F_{T/L}^{(a)}$
and $F_{T/L}^{(b)}$ corresponding to diagrams (a) and (b), respectively, 
are
\beq
F_{T/L}^{(a)} \, \vec{\epsilon}_{\lambda,\text{T/L}}\cdot\vec{S}_{M_J^\prime
  M_J}  =  \frac{3}{2}\, 
\bigg\langle M_J'\bigg|\frac{\vec{\epsilon}_{\lambda,\text{T/L}}\cdot (\vec \sigma_1
  + \vec \sigma_2) 
(\vec{\tau}_1\cdot\vec{\tau}_2-
\tau_1^z\tau_2^z)}{\vec{q}^{\;\prime2}}
\bigg| M_J\bigg\rangle_\psi~, 
\label{eq:valF2Na}
\eeq
and
\beq
F_{T/L}^{(b)} \, \vec{\epsilon}_{\lambda,\text{T/L}}\cdot\vec{S}_{M_J^\prime
  M_J} =  
 3 \, \bigg\langle M_J' \bigg|\,
\frac{    
(\vec{\tau}_1\cdot\vec{\tau}_2-\tau_1^z\tau_2^z)     
\big[(\vec{q}^{\;\prime}-\vec{k})\cdot (\vec \sigma_1 + \vec \sigma_2 ) \big] 
\big[\vec{\epsilon}_{\lambda,\text{T/L}}\cdot (\vec{q}^{\;\prime} -\vec{k}/2) \big]\, 
}
{\big[(\vec{q}^{\;\prime}-\vec{k})^2 +M_{\pi^+}^2 \big]
\vec{q}^{\;\prime2}} 
\bigg|M_J\bigg\rangle_\psi~,  
\label{eq:valF2Nb}
\eeq
where $\vec{q}^{\;\prime} = \vec{p}_{12} - \vec{p}_{12}^{\;\prime} +
\vec{k}/2$ is the momentum of the exchanged pion and 
 $\vec{p}_{12} =(\vec{k}_1-\vec{k}_2)/2$,
$\vec{p}_{12}^{\;\prime}=(\vec{k}_1^\prime-\vec{k}_2^\prime)/2$ 
are the initial and final Jacobi momenta
of nucleons 1 and 2, respectively. 
The integral for the 
form factors $F_{T/L}^{(a)}$ contains an integrable singularity which can
be removed by an appropriate variable transformation. Then, 
the form factors can 
be evaluated using Monte Carlo integration in the same way as the 
form factors for the single-nucleon contribution.
Our results for $F_{T/L}^{(a)}-F_{T/L}^{(b)}$ are given
in Table~\ref{tab:Fab}.
\renewcommand{\arraystretch}{1.3}
\begin{table}[t]
\begin{center}
\begin{tabular}{|c||c|c|}
\hline 
nucleus & $^3$He & $^3$H \\
\hline
$F_{T}^{(a)} -F_{T}^{(b)}$ [fm$^{-1}$]
&  $-29.3(2)(1)$ & $-29.7(2)(1)$ \\
$F_{L}^{(a)} -F_{L}^{(b)}$ [fm$^{-1}$]
&  $-22.9(2)(1)$ & $-23.2(1)(1)$ \\
\hline
\end{tabular}
 \caption{Numerical results for the form factors 
  $F_{T/L}^{(a)} -F_{T/L}^{(b)}$ parameterizing two-body contributions in units of fm$^{-1}$.
The first error is 
our estimation of the theoretical uncertainty resulting from the
truncation of the chiral expansion 
while the second one
   is the statistical error from the Monte Carlo integration.}
  \label{tab:Fab}
\end{center}
\end{table}
The first error is again the theory error estimated from the cutoff variation
in the chiral interaction as described above.
The second error is the statistical error from the Monte Carlo integration
which is about half the size of the theory error.

\subsection{Fourth order contributions}

\begin{figure}[t!]
        \begin{center}
        \includegraphics*[width=0.6\linewidth]{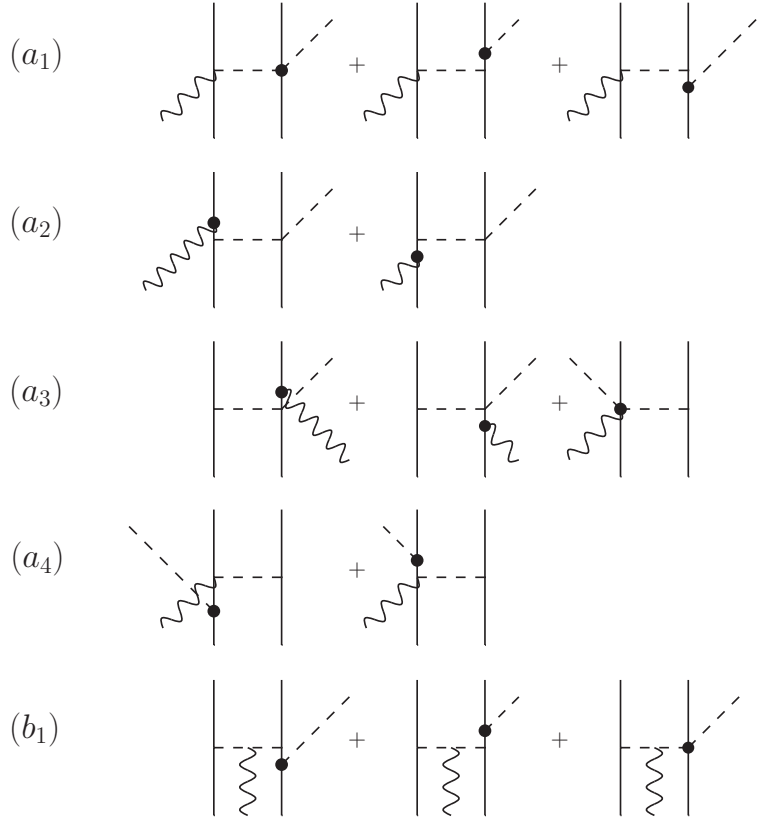}
        \end{center}
        \vspace{-0.2cm}
        \caption{Subleading static two-nucleon contributions to the 
         nuclear pion production matrix element at threshold. The
         filled circle denotes an insertion from the dimension two
         effective Lagrangian. For further notations, 
         see Fig.~\ref{fig:abterm}.}
        \label{fig:abterms}
        \end{figure}

\begin{figure}[h!]
        \begin{center}
        \includegraphics*[width=0.8\linewidth]{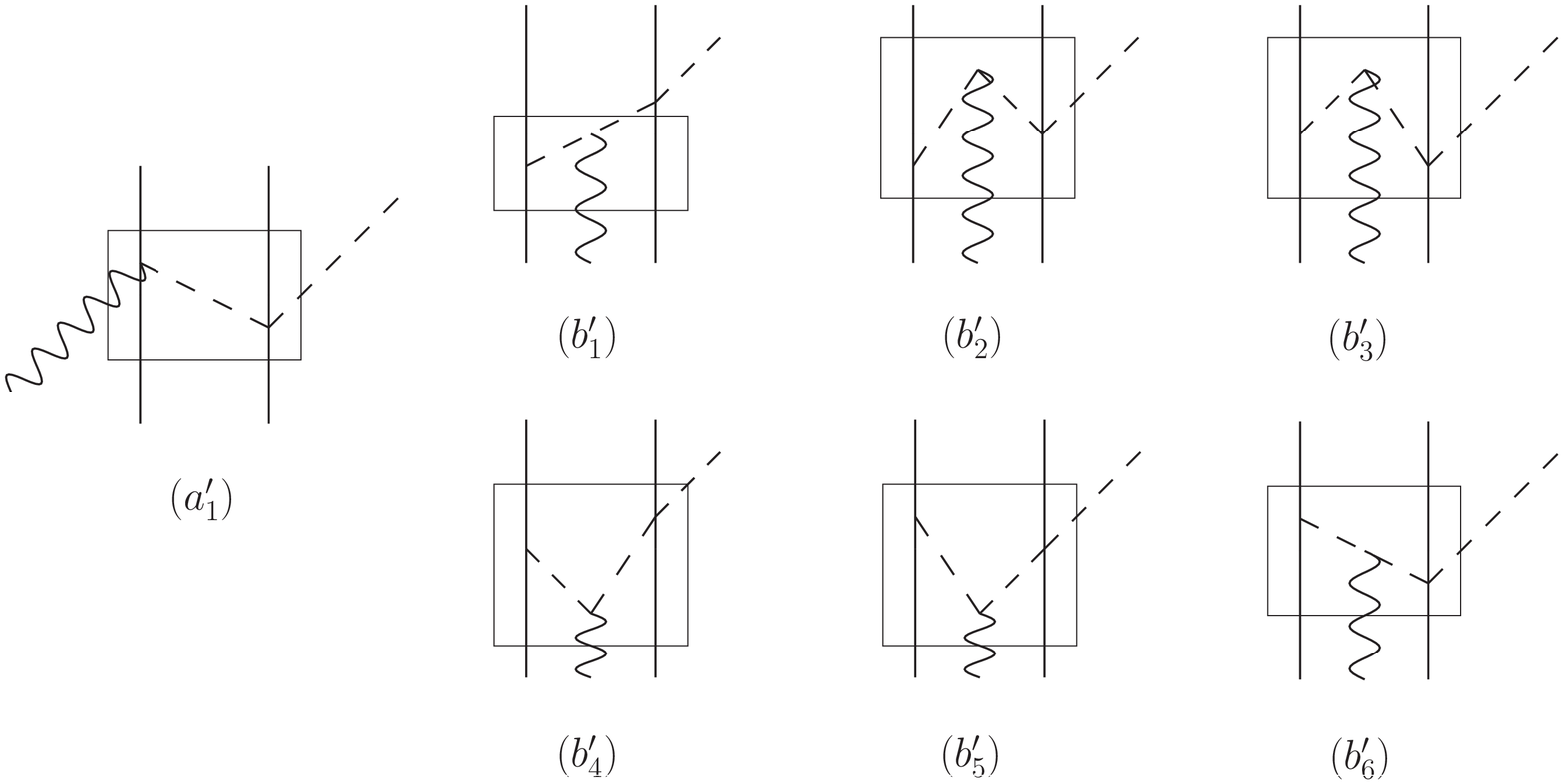}
        \end{center}
        \vspace{-0.2cm}
        \caption{Subleading recoil two-nucleon contributions to the 
         nuclear pion production matrix element at threshold in time-ordered
         perturbation theory. 
         The boxes indicate the regions where the two energy denominators
         to be expanded can appear
         Further notation as in Fig.~\ref{fig:abterm}.}
        \label{fig:abtermr}
        \end{figure}

Apart from the  fourth order corrections already included
in the single nucleon multipoles $E_{0+}$ and $L_{0+}$, there
are various fourth order corrections that arise due to the
presence of the other two nucleons in the tri-nucleon system
considered here. First, there are the boost corrections that
arise from the observation that within a nucleus the threshold
for pion production is shifted compared to the free threshold.
For light systems as considered here, this essentially induces
P-wave contributions as detailed in Sec.~\ref{sec:boost}.
Second, there are further corrections to the two-nucleon production operator 
displayed in Figs.~\ref{fig:abterms} and \ref{fig:abtermr}. 
%
%
These can be 
grouped in two categories,
namely the so-called static  and the so-called recoil corrections.
The static corrections are shown in Fig.~\ref{fig:abterms}.
They involve - as the leading 2N corrections do - static
propagators but one insertion from the dimension two chiral effective
pion-nucleon Lagrangian. The recoil corrections feature corrections
to the static propagators with only insertions from the leading
order (dimension one) chiral Lagrangian. These corrections are most
conveniently derived in time-ordered perturbation theory. The
corresponding diagrams are shown in Fig.~\ref{fig:abtermr}. The 
boxes indicate the regions where the two energy denominators
whose $1/m_N$-expansion generates these corrections can appear 
(see Ref.~\cite{Krebs:2003} for more details). 
These two types of corrections
will be discussed in Sec.~\ref{sec:stat} and Sec.~\ref{sec:recoil}, in
order.
%
%

\subsubsection{Boost corrections to the one-nucleon contributions}
\label{sec:boost}

The proton and neutron production amplitudes are calculated in the
$(N,\gamma)$ center-of-mass system (cms). The boost of a $(3N,\gamma)$-cms 
four-vector $p$ to the $(N,\gamma)$-cms four-vector $p^*$ has the general form
\begin{align}
\begin{pmatrix}
 p^{0*} \\ \vec{p}^{\,*}
\end{pmatrix}=
\begin{pmatrix}
 \gamma & -\gamma\vec{\beta}\\
 -\gamma\vec{\beta} & (\eins_3-P_{\vec{\beta}})+\gamma P_{\vec{\beta}}
\end{pmatrix}
\begin{pmatrix}
 p^0 \\ \vec{p}
\end{pmatrix}
=
\begin{pmatrix}
 \gamma (p^0-\vec{\beta}\cdot\vec{p}) \\ -\gamma\vec{\beta} p^0 + \vec{p}_\perp + \gamma\vec{p}_\parallel
\end{pmatrix},
\end{align}
where $P_{\vec{\beta}}$ is the projection operator onto the $\vec{\beta}$-direction, 
i.e. $P_{\vec{\beta}}\;\vec{x}=(\hat{\beta}\cdot\vec{x})\hat{\beta}$, 
$\vec{p}_\parallel=P_{\vec{\beta}}\;\vec{p}$ is the parallel part, 
and $\vec{p}_\perp=(1-P_{\vec{\beta}})\vec{p}$ the perpendicular part of 
$\vec{p}=\vec{p}_\parallel+\vec{p}_\perp$ with respect to $\vec{\beta}$. 
To determine $\vec{\beta}$, we consider $\vec{k}_1+\vec{k}$. 
In the $(N,\gamma)$-cms this combination has to vanish, i.e. $\vec{k}^{\,*}_1+\vec{k}^{\,*}=0$. 
We have
\begin{align}
\vec{k}_1^{\,*}+\vec{k}^{\,*}=\gamma\left(-\vec{\beta}
(k_1^0+k^0)+P_{\vec{\beta}}(\vec{k}_1+\vec{k})\right)+
(1-P_{\vec{\beta}})(\vec{k}_1+\vec{k})\stackrel{\mathrm{!}}=\vec{0}.
\end{align}
Because linearly independent (even orthogonal) vectors have to vanish separately, i.e.
\begin{align}
(1-P_{\vec{\beta}})(\vec{k}_1+\vec{k})&=\vec{0}~,\nonumber\\
-\vec{\beta}(k_1^0+k^0)+P_{\vec{\beta}}(\vec{k}_1+\vec{k})&=\vec{0}~,\nonumber
\end{align}
we conclude that
\begin{align}
\vec{\beta}=\frac{\vec{k}_1+\vec{k}}{k_1^0+k^0}=
\frac{\vec{k}_1^{\,\prime}+\vec{q}}{k_1^{\prime 0}+q^0}=\frac{\vec{p}_{12}^{\,\prime}-
\vec{p}_3^{\,\prime}/2+2\vec{q}/3}{\sqrt{(\vec{p}_{12}^{\,\prime}-\vec{p}_3^{\,\prime}/2-\vec{q}/3)^2+m_N^2}+q^0}.
\end{align}
For the definition of the Jacobi momenta used above, 
see Eqs.~(\ref{eq:valF2Na}, \ref{eq:valF2Nb}).
Near the static limit we have
\begin{align}
\vec{\beta} &=\frac{\vec{p}_{12}^{\,\prime}-\vec{p}_3^{\,\prime}/2+2\vec{q}/3}{m_N}\left(\sqrt{\frac{(\vec{p}_{12}^{\,\prime}-
\vec{p}_3^{\,\prime}/2-\vec{q}/3)^2}{m_N^2}+1}+\frac{\sqrt{\vec{q}^{\,2}+M_\pi^2}}{m_N}\right)^{-1}
\nonumber\\
&=\frac{\vec{p}_{12}^{\,\prime}-\vec{p}_3^{\,\prime}/2+2\vec{q}/3}{m_N}+\mathcal{O}\left( m_N^{-2}\right)
\xrightarrow{\text{threshold}}\frac{\vec{p}_{12}^{\,\prime}-\vec{p}_3^{\,\prime}/2}{m_N}+\mathcal{O}\left(m_N^{-2}\right),
\end{align}
and thus 
$\gamma  =
\left(1-\beta^2\right)^{-1/2}=1+\mathcal{O}\left(m_N^{-2}\right)$.  
Correspondingly, a general four-vector $p^\mu$ in the $(3N,\gamma)$-cms
transforms to the $(N,\gamma)$-cms as
\begin{align}
p^{0*} &=\gamma (p^0-\vec{\beta}\cdot\vec{p})=p^0-\frac{\vec{p}_{12}^{\,\prime}-
\vec{p}_3^{\,\prime}/2+2\vec{q}/3}{m_N}\cdot\vec{p}\xrightarrow{\text{threshold}}p^0-
\frac{\vec{p}_{12}^{\,\prime}-\vec{p}_3^{\,\prime}/2}{m_N}\cdot\vec{p},\\
\vec{p}^{\,*} &=-\gamma\vec{\beta} p^0 + \vec{p}_\perp + \gamma\vec{p}_\parallel=
\vec{p}-\frac{\vec{p}_{12}^{\,\prime}-\vec{p}_3^{\,\prime}/2+2\vec{q}/3}{m_N} p^0
\xrightarrow{\text{threshold}}\vec{p}-\frac{\vec{p}_{12}^{\,\prime}-\vec{p}_3^{\,\prime}/2}{m_N} p^0,
\end{align}
to first order in the inverse nucleon mass. At threshold, 
the boosted energy and momentum for the photon ($k^{0*}$, $\vec{k}^*$) and pion ($q^{0*}$, $\vec{q}^*$),
as well as the time- and space-like components of the
photon polarization vector ($\epsilon^{0*}_\lambda$, $\vec{\epsilon}^*_\lambda$)
are thus given by:
\begin{align}
k^{0*} &=  k^0 - \frac{(\vec{p}_{12}^{\,\prime}-\vec{p}_3^{\,\prime}/2)\cdot\vec{k}}{m_N}, 
&\vec{k}^*&=\vec{k} - \frac{k^0}{m_N}(\vec{p}_{12}^{\,\prime}-\vec{p}_3^{\,\prime}/2),\nonumber\\
q^{0*} &= q^0 - \frac{(\vec{p}_{12}^{\,\prime}-\vec{p}_3^{\,\prime}/2)\cdot\vec{q}}{m_N}=M_\pi, 
&\vec{q} \, ^*&=\vec{q} - \frac{q^0}{m_N}(\vec{p}_{12}^{\,\prime}-\vec{p}_3^{\,\prime}/2)=
-\frac{M_\pi}{m_N}(\vec{p}_{12}^{\,\prime}-\vec{p}_3^{\,\prime}/2),\nonumber\\
\epsilon^{0*}_\lambda 
&= \epsilon^0_\lambda - \frac{(\vec{p}_{12}^{\,\prime}-\vec{p}_3^{\,\prime}/2)\cdot\vec{\epsilon}_\lambda}{m_N}, 
&\vec{\epsilon}_\lambda^{\, *}
&=\vec{\epsilon}_\lambda - \frac{\epsilon^0_\lambda}{m_N}(\vec{p}_{12}^{\,\prime}-\vec{p}_3^{\,\prime}/2)=\vec{\epsilon}_\lambda.
\end{align}
Since $\epsilon^0_\lambda=0$, the polarization vector does not change except for the time component. 

To adopt the results for pion production off nucleons which were calculated in the $(N,\gamma)$-cms
for our calculation on the tri-nucleon, 
the pion momentum in the $(3N,\gamma)$-cms at threshold, $\vec{q}=\vec{0}$, is boosted to the 
$(N,\gamma)$-cms value $\vec{q}^{\,*}=-({M_\pi}/{m_N})(\vec{p}_{12}^{\,\prime}-\vec{p}_3^{\,\prime}/2)=:-\mu(\vec{p}_{12}^{\,\prime}-\vec{p}_3^{\,\prime}/2)$ 
above threshold. The corresponding P-wave contribution reads (using the
notation from Ref.~\cite{Bernard:1996bi})
\begin{align}
\mathcal{M}_\lambda &=
2i(\vec{\epsilon}_\lambda\cdot\vec{S})(\hat{q}^{\,*}\cdot\hat{k})P_1
+2i(\vec{\epsilon}_\lambda\cdot\hat{q}^{\,*})(\hat{k}\cdot\vec{S})P_2
+(\vec{\epsilon}_\lambda\cdot[\hat{q}^{\,*}\times\hat{k}])P_3 \nonumber\\
&+2i(\vec{\epsilon}_\lambda\cdot\hat{k})(\hat{k}\cdot\vec{S})(\hat{q}^{\,*}\cdot\hat{k})(P_4-P_5-P_1-P_2)
+2i(\vec{\epsilon}_\lambda\cdot\hat{k})(\hat{q}^{\,*}\cdot\vec{S})P_5\nonumber\\
&=:2i(\vec{\epsilon}_{\lambda,\text{T}}\cdot\vec{S})(\hat{q}^{\,*}\cdot\hat{k})P_1
+2i(\vec{\epsilon}_{\lambda,\text{T}}\cdot\hat{q}^{\,*})(\hat{k}\cdot\vec{S})P_2
+(\vec{\epsilon}_{\lambda,\text{T}}\cdot[\hat{q}^{\,*}\times\hat{k}])P_3\nonumber\\
&+2i(\vec{\epsilon}_{\lambda,\text{L}}\cdot\vec{S})(\hat{q}^{\,*}\cdot\hat{k})P_4
+2i(\vec{\epsilon}_{\lambda,\text{L}}\cdot\hat{k})(\hat{q}^{\,*}_{\text{T}}\cdot\vec{S})P_5. \nonumber
\end{align}
Close to threshold, the P-wave multipoles $P_i$ behave as 
$P_i\approx\bar{P_{i}}|\vec{q\,}^{\,*}|=\mu\bar{P_{i}}|\vec{p\,}_{12}^\prime-\vec{p}_3^{\,\prime}/2|$ with 
\beqa\label{eq:Pmult}
\bar{P_1^p} &=& +0.0187~\text{fm}^2~,  \quad \bar{P_3^p}=+0.0240~\text{fm}^2~,  
\quad \bar{P_4^p}=+0.0013~\text{fm}^2~,\nonumber\\
\bar{P_1^n} &=& +0.0134~\text{fm}^2~, \quad
\bar{P_3^n}=+0.0234~\text{fm}^2~,  \quad \bar{P_4^n}=+0.0003~\text{fm}^2~,
\eeqa
where the numerical values refer to the P-wave low-energy theorems for pion photo-
\cite{Bernard:1994gm} and electroproduction \cite{Bernard:1994dt}. Corrections
to these theorems are beyond the accuracy of our calculation.

In analogy to the S-wave case discussed above, we define the corresponding P-wave form factors
\begin{align}
\left(\vec{\epsilon}_{\lambda,\text{T}}\cdot\vec{S}_{M_J^\prime M_J}\right)F^{S/V}_1
&=\Braket{M_{J^\prime}|3\left(\vec{\epsilon}_{\lambda,\text{T}}\cdot\vec{S}_1\right)\left((\vec{p}_{12}^{\,
    \prime}-\vec{p}_3^{\,\prime}/2)\cdot\hat{k}\right) \,  X^{S/V}
|M_J}_{\psi},\nonumber \\
\left(\vec{\epsilon}_{\lambda,\text{T}}\cdot\vec{S}_{M_J^\prime M_J}\right)F^{S/V}_2
&=\Braket{M_{J^\prime}|3\Bigl(\vec{\epsilon}_{\lambda,\text{T}}\cdot(\vec{p}_{12}^{
    \, \prime}-\vec{p}_3^{\,\prime}/2)\Bigr)\left(\hat{k}\cdot\vec{S_1}\right) \,  X^{S/V}
|M_J}_{\psi},\nonumber \\
\left(\vec{\epsilon}_{\lambda,\text{T}}\cdot\vec{S}_{M_J^\prime M_J}\right)F^{S/V}_3
&=\Braket{M_{J^\prime}|-\frac{3}{2}i\left(\vec{\epsilon}_{\lambda,\text{T}}\cdot[(\vec{p}_{12}^{
    \, \prime}-\vec{p}_3^{\,\prime}/2)\times\hat{k}]\right) \,  X^{S/V}
|M_J}_{\psi},\nonumber \\
\left(\vec{\epsilon}_{\lambda,\text{L}}\cdot\vec{S}_{M_J^\prime M_J}\right)F^{S/V}_4
&=\Braket{M_{J^\prime}|3\left(\vec{\epsilon}_{\lambda,\text{L}}\cdot\vec{S}_1\right)\left((\vec{p}_{12}^{\,
    \prime}-\vec{p}_3^{\,\prime}/2)\cdot\hat{k}\right)
\,  X^{S/V}|M_J}_{\psi},\nonumber \\
\left(\vec{\epsilon}_{\lambda,\text{L}}\cdot\vec{S}_{M_J^\prime M_J}\right)F^{S/V}_5
&=\Braket{M_{J^\prime}|3\left(\vec{\epsilon}_{\lambda,\text{L}}\cdot\hat{k}\right)\left((\hat{p}_{12}^{\,
    \prime}-\vec{p}_3^{\,\prime}/2)_\text{T}\cdot\vec{S}_1\right)\,
  X^{S/V} |M_J}_{\psi}~,
\end{align}
where the spin and isospin operators refer to nucleon 1. 
Further, we have introduced the notation $X^{S} = 1$, $X^V =
\tau_1^z$.   
The contributions from the other nucleons are
accounted for by the overall factor of three.

In terms of these form factors, the P-wave contribution to the 3N-production amplitude
takes the form:
\begin{align}
E_{0+}^{1N, {\rm boost}}
\approx
-0.546 \mu\, \sum_{i=1}^3 \left(F_i^{S+V}\,\bar{P_i^p} + F_i^{S-V}\,\bar{P_i^n}\right)~,\\
L_{0+}^{1N, {\rm boost}}
\approx
-0.546 \mu\, \sum_{i=4}^5 \left(F_i^{S+V} \, \bar{P_i^p} + F_i^{S-V}\,\bar{P_i^n}\right)~,
\end{align}
where $F_i^{S\pm V}= {F_i^S\pm F_i^V}$.
These form factors are evaluated using the same Monte Carlo method as employed for the
S-waves. The numerical values are collected in Tab.~\ref{tab:FTLp}. Note that
$F_2$ and $F_5$ come out to be consistent with zero and therefore not
listed in the table. As before,
the proton contribution is dominant in $^3$H, whereas the neutron one
features prominently in $^3$He. 
\begin{table}[t]
\begin{center}
\begin{tabular}{|c||c|c|}
\hline 
nucleus & $^3$He  & $^3$H  \\
\hline
$F_{1}^{S+V}$ &  $+0.004(3)(1)$ & $+0.339(6)(1)$ \\ 
$F_{1}^{S-V}$ &  $+0.338(5)(1)$ & $+0.002(3)(1)$ \\
$F_{3}^{S+V}$ &  $-0.015(2)(0)$ & $-0.011(2)(0)$ \\
$F_{3}^{S-V}$ &  $-0.011(2)(0)$ & $-0.015(2)(0)$ \\
$F_{4}^{S+V}$ &  $-0.019(5)(4)$ & $+0.339(6)(4)$ \\
$F_{4}^{S-V}$ &  $+0.337(6)(4)$ & $-0.021(3)(4)$ \\
\hline
\end{tabular}
 \caption{Numerical results for the boost correction form factors $F_{i}^{S\pm
     V}$ in units of $[{\rm fm}^{-1}]$.
The first error is 
our estimation of the theoretical uncertainty resulting from the
truncation of the chiral expansion 
while the second one
   is the statistical error from the Monte Carlo integration.}
  \label{tab:FTLp}
\end{center}
\end{table}

Notice that in contrast to the single-nucleon corrections, we do not
need to employ a special treatment for boost corrections to 
the leading two-nucleon contributions. All $1/m_N$-corrections to the
leading three-body  contributions to the production operator needed in the
calculations are treated on the same footing as described in section \ref{sec:recoil}.

\subsubsection{Fourth order two-nucleon contributions}
\label{sec:stat}

Similar to the leading order  graphs
displayed in Fig.~\ref{fig:abterm},
one has additional contributions at NNLO, which are displayed as diagram
classes $a_1, a_2,a_3,a_4,$ and $b_1$ in Fig.~\ref{fig:abterms}. These involve the
insertion of one dimension-two operator from the chiral effective Lagrangian.
As in the case of the deuteron, only the operators $\sim 1/2m_N$, $\sim g_A/2m_N$
or $\sim \kappa_{p,n}$ contribute, while all contributions proportional to the
LECs $c_i$ vanish at threshold.  All these Feynman diagrams give corrections of the form 
$iT_{12}=\mathcal{N}\hat{\mathcal{O}}_{12}$ and involve multiples of 
the generic prefactor $\mathcal{N}=({eg_A m_N})/({2F_\pi^3})$. 
They lead to contributions of the form 
\begin{align}
\mathcal{M}&=2iK_{2N}^{q^4}\Braket{\hat{\mathcal{O}}_{12}+\hat{\mathcal{O}}_{21}}_\Psi
\end{align}
with
\begin{align}\label{eq:pref}
K_{2N}^{q^4}&=-\frac{1}{2}\frac{m_{3N}}{m_N}\frac{1}{8\pi(m_{3N}+M_\pi)}\frac{\mathcal{N}}{(2\pi)^3 2m_N}
= -K_{2N}\frac{1}{4m_N M_\pi}
\approx -0.0104 \,\text{fm}^3 \frac{10^{-3}}{M_{\pi^+}}
\end{align}
to account for the phase space and normalization. The prefactor of the 
2N contributions 
at order $q^4$, $K_{2N}^{q^4}$ is suppressed considerably  
compared to the corresponding prefactor at order $q^3$, $K_{2N}$.
This has to be kept in mind when we discuss the numerical results for the
fourth order corrections.

We are now in a position to evaluate the various classes of diagrams,
following the calculations for the deuteron from Refs.~\cite{Krebs:2004ir,Krebs:2003}.
The corrections to diagram $(a)$ in Fig.~\ref{fig:abterm} 
which are given in Fig.~\ref{fig:abterms}$(a_1)$-$(a_4)$ read
\begin{align}
iT^{NN,a_1}_{12}&=(1-2g_A^2)\frac{e m_N g_A}{2 F_\pi^3}\frac{\vec{\epsilon}_{\lambda,\text{T}}\cdot
\vec{\sigma}_{1}}{\vec{q}^{\;\prime2}}(\vec{\tau}_1\cdot\vec{\tau}_2-\tau_1^z\tau_2^z)
\,\left(\vec{q}^{\,\prime}\cdot\vec{q}^{\,\prime}+2\vec{q}^{\,\prime}\cdot\vec{p}_{12}^\prime\right)~,
\nonumber \\
iT^{NN,a_2}_{12}&=\frac{e m_N g_A}{F_\pi^3}\frac{\vec{\epsilon}_{\lambda,\text{T}}\cdot\left(
(\vec{q}^{\;\prime}+2\vec{p}_{12}^{\;\prime}-\vec{k})\,\vec{\sigma}_{1}\cdot\vec{q}^{\;\prime}
+i[\vec{q}^{\;\prime}\times\vec{k}](1+\kappa_V)\right)}{\vec{q}^{\;\prime2}}
(\vec{\tau}_1\cdot\vec{\tau}_2-\tau_1^z\tau_2^z)~,
\nonumber \\
iT^{NN,a_3}_{12}&=\frac{e m_N g_A}{F_\pi^3}\frac{\vec{\epsilon}_{\lambda,\text{T}}\cdot\left(\vec{q}^{\;\prime}
+2\vec{p}_{12}^{\;\prime}-\vec{k}+i[\vec{\sigma}_1\times\vec{k}](1+\kappa_V)\right)
\vec{\sigma}_{2}\cdot\vec{q}^{\;\prime}}{\vec{q}^{\;\prime2}+M^2_{\pi^+}}
(\vec{\tau}_1\cdot\vec{\tau}_2-\tau_1^z\tau_2^z)~,
\nonumber \\
  \label{eq:T2Na4}
iT^{NN,a_4}_{12}&=g_A^2\frac{e m_N g_A}{F_\pi^3}\frac{\vec{\epsilon}_{\lambda,\text{T}}\cdot
\left(-(\vec{q}^{\;\prime}+2\vec{p}_{12}^{\;\prime}-\vec{k})+i[\vec{\sigma}_1\times
(\vec{q}^{\;\prime}-\vec{k})]\right)\vec{\sigma}_{2}\cdot\vec{q}^{\;\prime}}{\vec{q}^{\;\prime2}+M^2_{\pi^+}}
(\vec{\tau}_1\cdot\vec{\tau}_2-\tau_1^z\tau_2^z)~.
\end{align}
The corrections to diagram $(b)$ in Fig.~\ref{fig:abterm} 
which are given in Fig.~\ref{fig:abterms}$(b_1)$ read
\begin{align}
  \label{eq:T2Nb1}
iT^{NN,b_1}_{12} =-\left(1-2g_A^2\right)\frac{e m_N g_A}{2
  F_\pi^3}\frac{\vec{\sigma}_1 \cdot \vec{q}^{\,\prime\prime}\;
  \vec{\epsilon}_{\lambda,\text{T}}\cdot(\vec{q}^{\,\prime\prime}+\vec{q}^{\,\prime})}{\left(\vec{q}^{\,\prime\prime}\cdot
\vec{q}^{\,\prime\prime}+M^2_{\pi^+}\right)\left(\vec{q}^{\,\prime}\cdot\vec{q}^{\,\prime}\right)}(\vec{\tau}_1\cdot\vec{\tau}_2-\tau_1^z\tau_2^z)\,
\left(\vec{q}^{\,\prime}\cdot\vec{q}^{\,\prime}+2\vec{q}^{\,\prime}\cdot\vec{p}_{12}^\prime\right).
\end{align}
with $\vec{q}^{\,\prime}$ as before and $\vec{q}^{\,\prime\prime} =
\vec{q}^{\, \prime} - \vec{k}$. 

\subsubsection{{\boldmath$1/m_N$}-corrections to leading two-nucleon contributions}
\label{sec:recoil}

Furthermore, there are $1/m_N$-corrections to leading two-nucleon
diagrams some of which are shown in Fig.~\ref{fig:abterm}. 
They have been worked out within the $Q$-box framework in
Ref.~\cite{Krebs:2003}.  
The corresponding diagrams $a'_1$, $b'_1$,  $b'_2$,
$b'_3$, $b'_4$, $b'_5$, and $b'_6$  are displayed in Fig.~\ref{fig:abtermr}. The 
boxes indicate the regions where the two energy denominators
whose $1/m_N$-expansion generates the corrections can appear 
(see Ref.~\cite{Krebs:2003} for more details). 

For the expressions of the various diagrams, we use the following
abbreviations: 
\beq\omega^\prime=\sqrt{\vec{q}^{\;\prime2}+M^2_{\pi^+}}\quad {\rm  and} \quad 
\omega^{\prime\prime}=\sqrt{\vec{q}^{\;\prime\prime2}+M^2_{\pi^+}}~.
\eeq
With that, the correction to diagram $(a)$ in Fig.~\ref{fig:abterm} 
which are given in Fig.~\ref{fig:abtermr}$(a'_1)$ read
\begin{align}
  \label{eq:T2Na1p}
iT^{NN,a^\prime_1}_{12}&=\frac{e m_N g_A}{8 F_\pi^3}\frac{\vec{\epsilon}_{\lambda,\text{T}}\cdot\vec{\sigma}_{1}(\omega^\prime-M_\pi)}{\omega^\prime(\omega^\prime+M_\pi)^2}\vec{k}\cdot(-2\vec{q}^{\,\prime}-2\vec{p}_{12}^{\;\prime}+\vec{k}) (\vec{\tau}_1\cdot\vec{\tau}_2-\tau_1^z\tau_2^z)~,
\end{align}
whereas the various corrections to diagram $(b)$ in Fig.~\ref{fig:abterm} 
which are given in Fig.~\ref{fig:abtermr}$(b'_1)$-$(b'_6)$ 
take the form
\begin{align}
iT^{NN,b^\prime_1}_{12}&=\frac{e m_N g_A}{16 F_\pi^3}\frac{\vec{\sigma}_1 \cdot \vec{q}^{\,\prime\prime} 
\vec{\epsilon}_{\lambda,\text{T}}\cdot(\vec{q}^{\,\prime\prime}+\vec{q}^{\,\prime})(\omega^\prime+M_\pi)}
{\omega^\prime\omega^{\prime\prime 3}}(\vec{\tau}_1\cdot\vec{\tau}_2-\tau_1^z\tau_2^z)\nonumber\\
&\times\left(\frac{-\vec{q}^{\,\prime}\cdot\vec{q}^{\,\prime}-\vec{q}^{\,\prime}\cdot\vec{p}_{12}^\prime
+2(\vec{q}^{\,\prime}+\vec{p}_{12}^\prime)\cdot\vec{k}-\vec{k}^2}{\omega^\prime-M_\pi}
+\frac{\vec{q}^{\,\prime}\cdot\vec{q}^{\,\prime}+\vec{q}^{\,\prime}\cdot\vec{p}_{12}^\prime}
{\omega^\prime-\omega^{\prime\prime}-M_\pi}\right)~,
\nonumber
\end{align}
\begin{align}
iT^{NN,b^\prime_2}_{12}&=\frac{e m_N g_A}{16 F_\pi^3}\frac{\vec{\sigma}_1 \cdot \vec{q}^{\,\prime\prime} 
\vec{\epsilon}_{\lambda,\text{T}}\cdot(\vec{q}^{\,\prime\prime}+\vec{q}^{\,\prime})(\omega^\prime-M_\pi)}{\omega^\prime
\omega^{\prime\prime 3}(\omega^\prime+\omega^{\prime\prime}+M_\pi)^2}
(\vec{\tau}_1\cdot\vec{\tau}_2-\tau_1^z\tau_2^z)\nonumber\\
&\times\left(2\omega^{\prime\prime}\left\{\vec{q}^{\,\prime}\cdot\vec{q}^{\,\prime}
+\vec{q}^{\,\prime}\cdot\vec{p}_{12}^\prime-2(\vec{q}^{\,\prime}+\vec{p}_{12}^\prime)\cdot\vec{k}
+\vec{k}^2\right\}+(\omega^\prime+M_\pi)\left\{-2(\vec{q}^{\,\prime}+\vec{p}_{12}^\prime)\cdot\vec{k}+
\vec{k}^2\right\}\right)~,
\nonumber \\
iT^{NN,b^\prime_3}_{12}&=\frac{e m_N g_A}{16 F_\pi^3}\frac{\vec{\sigma}_1 \cdot \vec{q}^{\,\prime\prime} 
\vec{\epsilon}_{\lambda,\text{T}}\cdot(\vec{q}^{\,\prime\prime}+\vec{q}^{\,\prime})(\omega^\prime-M_\pi)}{\omega^\prime
\omega^{\prime\prime 3}(\omega^\prime+\omega^{\prime\prime}+M_\pi)^2(\omega^\prime+\omega^{\prime\prime})^2}
(\vec{\tau}_1\cdot\vec{\tau}_2-\tau_1^z\tau_2^z)\nonumber\\
&\times\left(\omega^{\prime\prime}\left\{2(\vec{q}^{\,\prime}+\vec{p}_{12}^\prime)\cdot\vec{k}-\vec{k}^2\right\}
+2(\omega^\prime+M_\pi)\left\{\vec{q}^{\,\prime}\cdot\vec{q}^{\,\prime}+\vec{q}^{\,\prime}\cdot
\vec{p}_{12}^\prime\right\}\right)~,
\nonumber \\
iT^{NN,b^\prime_4}_{12}&=\frac{e m_N g_A}{16 F_\pi^3}\frac{\vec{\sigma}_1 \cdot \vec{q}^{\,\prime\prime} 
\vec{\epsilon}_{\lambda,\text{T}}\cdot(\vec{q}^{\,\prime\prime}+\vec{q}^{\,\prime})(\omega^\prime+M_\pi)}{\omega^\prime
\omega^{\prime\prime 3}(\omega^\prime+\omega^{\prime\prime}+M_\pi)^2(\omega^\prime+\omega^{\prime\prime})^2}
(\vec{\tau}_1\cdot\vec{\tau}_2-\tau_1^z\tau_2^z)\nonumber\\
&\times\left(\omega^{\prime\prime}\left\{\vec{q}^{\,\prime}\cdot\vec{q}^{\,\prime}+\vec{q}^{\,\prime}\cdot
\vec{p}_{12}^\prime-2(\vec{q}^{\,\prime}+\vec{p}_{12}^\prime)\cdot\vec{k}+\vec{k}^2\right\}
-(\omega^\prime-M_\pi)\left\{\vec{q}^{\,\prime}\cdot\vec{q}^{\,\prime}+
\vec{q}^{\,\prime}\cdot\vec{p}_{12}^\prime\right\}\right)~,
\nonumber \\ 
iT^{NN,b^\prime_5}_{12}&=\frac{e m_N g_A}{16 F_\pi^3}\frac{\vec{\sigma}_1 \cdot \vec{q}^{\,\prime\prime} 
\vec{\epsilon}_{\lambda,\text{T}}\cdot(\vec{q}^{\,\prime\prime}+\vec{q}^{\,\prime})(\omega^\prime+M_\pi)}{\omega^\prime
\omega^{\prime\prime 3}(\omega^\prime+\omega^{\prime\prime}-M_\pi)^2}
(\vec{\tau}_1\cdot\vec{\tau}_2-\tau_1^z\tau_2^z)\nonumber\\
&\times\left(2\omega^{\prime\prime}\left\{\vec{q}^{\,\prime}\cdot\vec{q}^{\,\prime}+
\vec{q}^{\,\prime}\cdot\vec{p}_{12}^\prime-2(\vec{q}^{\,\prime}+\vec{p}_{12}^\prime)\cdot\vec{k}+
\vec{k}^2\right\}+(\omega^\prime-M_\pi)\left\{-2(\vec{q}^{\,\prime}+\vec{p}_{12}^\prime)\cdot\vec{k}+
\vec{k}^2\right\}\right)~,
\nonumber \\
  \label{eq:T2Nb6p}
iT^{NN,b^\prime_6}_{12}&=\frac{e m_N g_A}{16 F_\pi^3}\frac{\vec{\sigma}_1 \cdot
  \vec{q}^{\,\prime\prime}
  \vec{\epsilon}_{\lambda,\text{T}}\cdot(\vec{q}^{\,\prime\prime}+\vec{q}^{\,\prime})(\omega^\prime-M_\pi)}
{\omega^\prime\omega^{\prime\prime 3}(\omega^\prime+M_\pi)^2}
(\vec{\tau}_1\cdot\vec{\tau}_2-\tau_1^z\tau_2^z)\nonumber\\
&\times \left((\omega^\prime+\omega^{\prime\prime}+M_\pi)\left\{2(\vec{q}^{\,\prime}+
\vec{p}_{12}^\prime)\cdot\vec{k}-\vec{k}^2\right\}\right).
\end{align}

\section{Results and discussion}

We are now in a position to evaluate the
nuclear S-wave multipoles. They are given as the
sum of the one- and two-nucleon contributions given
in the previous section,
\beq
E_{0+} = E_{0+}^{\rm 1N}+E_{0+}^{\rm 2N}~,\quad
L_{0+} = L_{0+}^{\rm 1N}+L_{0+}^{\rm 2N}~.
\eeq
Combining the leading and subleading corrections to the
two-nucleon production operators discussed above with the
subleading chiral perturbation theory results for the 
single-nucleon multipoles at ${\cal O}(p^4)$
\cite{Bernard:1994gm,Bernard:2001gz,Bernard:2005dj}
\beqa
E_{0+}^{\pi^0 p} &=& -1.16 \times 10^{-3}/M_{\pi^+}~,\quad 
E_{0+}^{\pi^0 n}  =  +2.13 \times 10^{-3}/M_{\pi^+}~,\nonumber\\
L_{0+}^{\pi^0 p} &=& -1.35 \times 10^{-3}/M_{\pi^+}~,\quad
L_{0+}^{\pi^0 n}  =  -2.41 \times 10^{-3}/M_{\pi^+}~,
\label{eq:1Nmultipoles}
\eeqa
we obtain for the threshold multipoles on $^3$He and on $^3$H
the values listed in Table~\ref{tab:multip}. For
$E_{0+}^{\pi^0 p}$ we took an average of 
Refs.~\cite{Bernard:2001gz,Bernard:2005dj}. The neutron amplitude
uses the updated LECs from \cite{Bernard:2001gz} based on the formalism 
from \cite{Bernard:2001gz} but is not explicitely given in that
paper. Note that the values 
for the single nucleon multipoles
in Eq.~(\ref{eq:1Nmultipoles}) are consistent with the 
unpolarized data
of \cite{Schmidt:2001vg} and a recent calculation in the chiral 
unitary approach \cite{Gasparyan:2010xz}. The
extractions of the proton S-wave photoproduction multipoles based
on CHPT using various approximations 
show a 5\% uncertainty \cite{FernandezRamirez:2009jb}.
Consequently, we assign a 5\% error to the single-nucleon 
multipoles.\footnote{We note that it is misleading to estimate the
theory uncertainty from comparing third and fourth order results 
due to the abnormally large contribution of the triangle diagram.
The theory uncertainty has therefore been estimated from a comparison
of fitting various data sets available and using variations of the ChPT 
amplitude that account for unitarity exactly.}

We remark that in the heavy baryon calculations of 
Refs.~\cite{Bernard:1994gm,Bernard:2001gz}, the physical pion masses
have been used in the kinematics. Moreover, the physical pion masses
were used in evaluating the corresponding loop diagrams. 
Thus the production as well as
the second threshold due to the $\pi^+ n$ intermediate state are correctly
accounted for. The same approach is used here. Therefore, the
multipole values in Eq. (\ref{eq:1Nmultipoles}) can indeed be tested in 
pion production off the tri-nucleon.

\begin{table}[t]
\centering
\begin{tabular}{|c||c|c|c|c|c|c|}
\hline 
$^3$He & 1N ($q^4$) & 2N ($q^3$) & 1N-boost & 2N-static ($q^4$) & 2N-recoil ($q^4$) & total \\
\hline
$E_{0+}$ [${10^{-3}}/{M_{\pi^+}}$]
& +1.71(4)(9) & $-$3.95(3) & $-$0.23(1) & $-$0.02(0)(1) & +0.01(2)(1) & $-$2.48(11)\\  
$L_{0+}$ [${10^{-3}}/{M_{\pi^+}}$]
& $-$1.89(4)(9) & $-$3.09(2) & $-$0.00(0) & $-$0.07(1)(1) & +0.07(7)(0) & $-$4.98(12)\\
\hline  
$^3$H & 1N ($q^4$) & 2N ($q^3$) & 1N-boost & 2N-static ($q^4$) & 2N-recoil ($q^4$) & total \\
\hline
$E_{0+}$ [${10^{-3}}/{M_{\pi^+}}$] 
& $-$0.93(3)(5) & $-$4.01(3) & $-$0.35(1) & $-$0.02(1)(1) & +0.01(2)(0) & $-$5.28(7)\\  
$L_{0+}$ [${10^{-3}}/{M_{\pi^+}}$]
& $-$0.99(4)(5) & $-$3.13(1) & $-$0.02(0) & $-$0.07(0)(1) & +0.07(7)(0) & $-$4.14(10)\\
\hline
\end{tabular}
\caption{Numerical results for the 3N 
multipoles.  
The first error is 
our estimation of the theoretical uncertainty resulting from the
truncation of the chiral expansion 
while the second one
   is the statistical error from the Monte Carlo integration
  (with the exception of the $1N$ contributions as explained in the text).
For the total result only the combined error is given.
\label{tab:multip}}
\end{table}
In Table~\ref{tab:multip}, adding the first two columns gives the leading one-loop
result from Ref.~\cite{Lenkewitz:2011jd}.
The fourth order corrections are given separately for the boost of the
single nucleon terms (third column) as well as the contributions
described in sections \ref{sec:stat} and \ref{sec:recoil} and referred
to as static and recoil, respectively. 
The complete one-loop result
can be found in the sixth column. The first error given is an estimate 
of the theory error from higher orders in chiral EFT, the
second error is the statistical error from the Monte Carlo
integration. 
Notice that the statistical error is negligible compared to theory error.
The 5\% error from the single-nucleon amplitudes discussed above 
is not included in the numbers for the theory error, but appears as
the second error of the single nucleon contribution in the table.
For the total result only the combined error is given.
Overall, we find that these fourth order corrections
for the electric dipole amplitude $E_{0+}$ for both tri-nuclear
systems come out to be very small, much smaller than in case of the
deuteron. This can be, in part, traced back to the smaller
values of the various form factors (for the boost corrections)
and also  to the small prefactor $K_{2N}^{q4}$,
cf. Eq.~(\ref{eq:pref}), for the two-nucleon contributions. 
For the longitudinal amplitude $L_{0+}$, the sum of the fourth order corrections
is consistent with zero within the uncertainties. This can be understood
as follows: First, the boost corrections are proportional to the P-wave
multipole $P_4$, which is 
much smaller than the corresponding
multipoles $P_1, P_3$ that appear in the electric dipole amplitude,
cf. Eq.~(\ref{eq:Pmult}). Second, there are almost perfect cancellations
between the static and the recoil contributions for both tri-nucleon
systems. These cancellations are accidental in the sense that they
can not be traced back to any symmetry or small prefactor.
In summary, we find that the chiral expansion for the S-wave multipoles
at threshold converges fast and that the largest uncertainty remains in
the single nucleon production amplitudes. We also remark that the 
fourth order corrections to the electric dipole amplitudes of the
tri-nucleon systems are sizably smaller than for the deuteron.

\begin{figure}[t]
        \begin{center}
        \includegraphics*[width=0.65\linewidth]{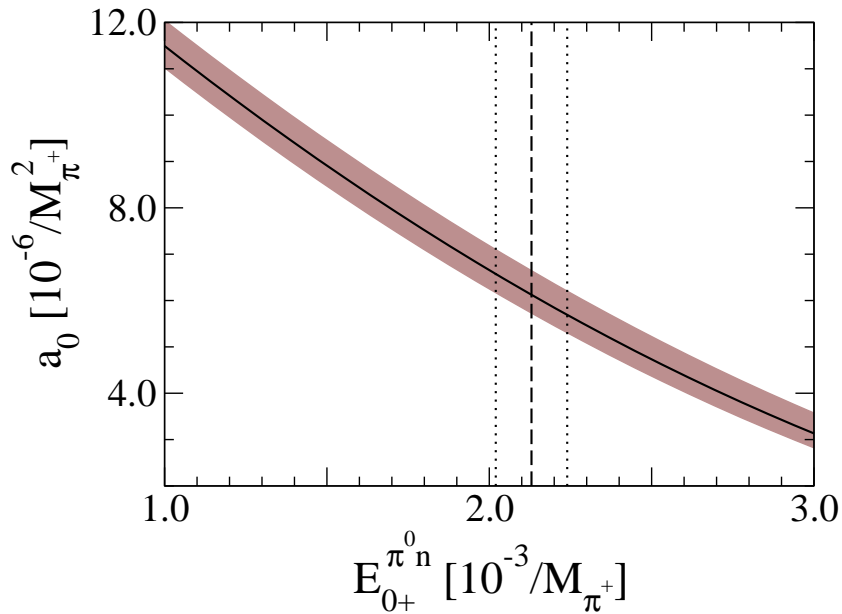}
        \end{center}
        \vspace{-0.2cm}
        \caption{ Sensitivity of $a_0$ for $^3$He in units of
        $ 10^{-6}/M^2_{\pi^+}$
        to the single-neutron multipole $E_{0+}^{\pi^0 n}$
        in units of $ 10^{-3}/M_{\pi^+}$. The vertical dashed line
        gives the CHPT prediction for $E_{0+}^{\pi^0 n}$ and the
        vertical dotted lines indicate the 5\% error in the prediction.
        The shaded band indicates the theory error estimated
        from the cutoff variation and a 5\% error in $E_{0+}^{\pi^0 p}$
        as described in the text.} 
        \label{fig:E0psq}
        \end{figure}

Now, let us concentrate on photoproduction. The threshold S-wave cross section for pion photoproduction $a_0$ is given by
\beq
a_0 =\left.\frac{|\vec{k}\,|}{|\vec{q}\,|}
\frac{d\sigma}{d\Omega}\right|_{\vec{q}=0}=\left|E_{0+}\right|^2~.
\eeq
In Fig.~\ref{fig:E0psq}, we illustrate the sensitivity of $a_0$ to the single-neutron multipole $E_{0+}^{\pi^0 n}$. 
The shaded band indicates the theory error estimated
from the cutoff variation as described above
and a 5\% error in $E_{0+}^{\pi^0 p}$.  As shown above,
the uncertainties related to the nuclear
effects are of the order of one percent, i.e. completely
negligible. So our estimate of a 10\% uncertainty of the
2N contributions in Ref.~\cite{Lenkewitz:2011jd}  driven by 
the analogy to the deuteron case~\cite{Beane:1997iv} 
turns out to be much too conservative.
The vertical dashed line indicates the CHPT prediction $E_{0+}^{\pi^0 n}= 2.13 \times 10^{-3}/M_{\pi^+}$.
Changing this value by $\pm20\%$ leads to changes in $a_0$ of about $\pm30\%$.
Thus, the $^3$He nucleus  is
a very promising target to test
the CHPT prediction for $E_{0+}^{\pi^0 n}$. On the contrary, neutral pion production on $^3$H is rather insensitive to $E_{0+}^{\pi^0 n}$:
a variation of $E_{0+}^{\pi^0 n}$ 
in the range $0 \ldots 3$ (in units of $10^{-3}/M_{\pi^+}$) changes  $a_0$
only by $1\%$.

Next we compare our predictions with the available data. 
The consistency of the CHPT prediction for the single-neutron 
multipole with the measured S-wave threshold amplitude on the
deuteron from Saclay and Saskatoon
is well established, see Refs.~\cite{Beane:1997iv,Bernard:2007zu}.
The reanalyzed measurement of the S-wave amplitude for $^3$He
at Saclay gives
$E_3 = (-3.5\pm 0.3)\times 10^{-3}/M_{\pi^+}$ \cite{Argan:1980zz,Argan:1987dm},
which is related to $a_0$ according to 
\beq
\left|E_{0+}\right|^2 = |E_3|^2 \left|\frac{F_T^{S-V}}{2} \right|^2 \left(
\frac{1+M_\pi/m_N}{1+M_\pi/3m_N}\right)^2~.
\eeq
Here, we have approximated the $A=3$ body form factor ${\cal F}_A$
of Argan et al.~\cite{Argan:1987dm} by the numerically dominant form factor
$F_{T}^{S-V}$ 
for $^3$He, cf.~Tab.~\ref{tab:FTL}. This results in 
\beq
E_{0+} = (-2.8 \pm 0.2) \times 10^{-3}/M_{\pi^+}~,
\label{eq:argan}
\eeq
assuming the same sign as for our $^3$He prediction in
Table~\ref{tab:multip}. In magnitude,
the extracted value is about 25\% above 
the leading one-loop prediction $E_{0+} = -2.24 (11) \times 10^{-3}/M_{\pi^+}$.
The discrepancy is reduced for the fourth order result
$E_{0+} = -2.48 (11) \times 10^{-3}/M_{\pi^+}~,$ which is about 12\% below the 
experimental value in Eq.~(\ref{eq:argan}). Taking into account the
errors, our fourth order result is consistent with the experiment of Argan 
et al.~\cite{Argan:1987dm}. Given the
model-de\-pen\-dence that is inherent to the analysis of Ref.~\cite{Argan:1987dm},
it is obvious that a more precise measurement using  CW beams and modern
detectors is very much called for.
Our calculation establishes
a model independent connection between the trinucleon multipole measured
in such an experiment and the single neutron multipole to be extracted.

\section{Summary and outlook}

In this paper, we have presented a calculation of the subleading
two-nucleon operators for threshold neutral pion photo- 
and electroproduction off the tri-nucleon systems $^3$H and  $^3$He.
This completes the one-loop calculation at order ${\cal O}(q^4)$
in the standard heavy baryon counting.
The single nucleon contributions to that order and the leading 
third order two-nucleon contributions were already evaluated in 
the corresponding letter \cite{Lenkewitz:2011jd}.
The production operator was evaluated  in the framework of 
chiral nuclear effective field theory, in line with the earlier
calculations for production off the deuteron 
\cite{Beane:1995cb,Beane:1997iv,Krebs:2004ir}.
To this order, it gets both
one- and two-body contributions. Here, we have given explicit expressions
for the fourth order two-nucleon  contributions stemming
from boost corrections for pion production off a single nucleon (such contributions only arise
from the fact that in a nucleus the threshold for pion production is
lowered as compared to the nucleon case), from static 2N contributions
with one insertion from the dimension two chiral pion-nucleon Lagrangian and from
recoil corrections to the pion- and nucleon propagators. 
We used the chiral wave functions of
Refs.~\cite{Epelbaum:2003gr,Epelbaum:2003xx} to calculate 
the S-wave 3N multipoles $E_{0+}$ and $L_{0+}$.
These wave functions are consistent with the  pion production
operator.

We have shown
that all corrections at fourth order in the standard heavy baryon counting
are very small, a few percent for the tri-nucleon
electric dipole amplitudes and essentially vanishing for the corresponding
longitudinal amplitudes. This suppression can be explained by very small
boost correction and an accidental cancellation between the static and
the recoil contibutions.  We remark that these corrections are sizably
smaller than in the deuteron case~\cite{Beane:1997iv}.
However, we note that the the role of nucleon recoil,
accounting for the new scale $\chi=\sqrt{M_\pi m_N}\simeq 340$ MeV,
needs to be investigated in more detail in
view of the findings of Refs.~\cite{Rekalo:2001he,Baru:2004kw}.

The theoretical uncertainty  associated with 
the cutoff variation in the
employed wave functions appears to be small (of the order of 
$3\%$).  The dominant theoretical error at this order stems
from the threshold pion production amplitude off the proton and the
neutron, which is estimated to be about 5\%. 
Consequently, we have explored the possibility to extract 
 the elementary neutron multipole $E_{0+}^{\pi^0 n}$ from a neutral
 pion photoproduction measurement off $^3$He. We found indeed a large
 sensitivity of the $E_{0+}$ amplitude to $E_{0+}^{\pi^0 n}$. 
Given the 
very small uncertainty of the nuclear corrections 
as shown here,
$^3$He
appears to be a promising target to test the counterintuitive CHPT prediction
for  $E_{0+}^{\pi^0 n}$~\cite{Bernard:1994gm,Bernard:2001gz}.

We have also shown that our prediction for the $^3$He S-wave multipole
$E_{0+}$ is roughly consistent with the  value deduced from the old
Saclay measurement of the threshold cross section \cite{Argan:1987dm}.
A new measurement using modern technology and better
methods to deal with  few-body dynamics is urgent\-ly called for. 
The rapid energy dependence of $E_{0+}^{\pi^0 n}$ due to the close by 
charged pion production threshold presents a challenge for its 
experimental extraction. An calculation of pion production above threshold
would be valuable in this context.

There are many other natural extensions of this work. They include investigating
higher orders, the extension to virtual photons
and pion electroproduction, production of charged pions, and 
considering heavier nuclear targets such as $^4$He.
Further work in these directions is in progress. 

\bigskip\bigskip

\noindent{\large {\bf Acknowledgements}}

\smallskip\noindent
We thank Andreas Nogga for providing us with the chiral 3N 
wave functions and Hermann Krebs for discussions.
Financial support by the Deutsche Forschungsgemeinschaft 
(SFB/TR 16, ``Subnuclear 
Structure of Matter''),
by the European Community Research Infrastructure Integrating Activity 
``Study of Strongly Interacting Matter''
(acronym HadronPhysics3, Grant A\-gree\-ment n.~283286) under the 7th
Framework Programme of the EU and by 
the European Research Council (acronym NuclearEFT, ERC-2010-StG 259218)
is gratefully
acknowled\-ged.



\begin{thebibliography}{99}

\bibitem{Phillips:2009af}
  D.~R.~Phillips,
  J.\ Phys.\ G {\bf 36} (2009) 104004
  [arXiv:0903.4439 [nucl-th]].


\bibitem{Bernard:2007zu}
  V.~Bernard,
  Prog.\ Part.\ Nucl.\ Phys.\  {\bf 60} (2008) 82
  [arXiv:0706.0312 [hep-ph]].


\bibitem{Bernard:1994gm}   
  V.~Bernard, N.~Kaiser and U.-G.~Mei{\ss}ner,
  Z.\ Phys.\  C {\bf 70} (1996) 483
  [arXiv:hep-ph/9411287].

\bibitem{Bernard:2001gz}
  V.~Bernard, N.~Kaiser and U.-G.~Mei{\ss}ner,
  Eur.\ Phys.\ J.\  A {\bf 11} (2001) 209
  [arXiv:hep-ph/0102066].

\bibitem{Beane:1995cb}
  S.~R.~Beane, C.~Y.~Lee and U.~van Kolck,
  Phys.\ Rev.\  C {\bf 52} (1995) 2914
  [arXiv:nucl-th/9506017].

\bibitem{Beane:1997iv}
  S.~R.~Beane, V.~Bernard, T.-S.~H.~Lee, U.-G.~Mei{\ss}ner and U.~van Kolck,
  Nucl.\ Phys.\  A {\bf 618} (1997) 381
  [arXiv:hep-ph/9702226].

\bibitem{Krebs:2004ir}
  H.~Krebs, V.~Bernard and U.-G.~Mei{\ss}ner,
  Eur.\ Phys.\ J.\  A {\bf 22} (2004) 503
  [arXiv:nucl-th/0405006].

\bibitem{Epelbaum:2008ga}
  E.~Epelbaum, H.-W.~Hammer and U.-G.~Mei{\ss}ner,
  Rev.\ Mod.\ Phys.\  {\bf 81} (2009) 1773
  [arXiv:0811.1338 [nucl-th]].

\bibitem{Machleidt:2011zz}
  R.~Machleidt and D.~R.~Entem,
  Phys.\ Rept.\  {\bf 503} (2011) 1
  [arXiv:1105.2919 [nucl-th]].

\bibitem{Lenkewitz:2011jd}
  M.~Lenkewitz, E.~Epelbaum, H.-W.~Hammer and U.-G.~Mei\ss ner,
  Phys.\ Lett.\ B {\bf 700} (2011) 365
  [arXiv:1103.3400 [nucl-th]].


\bibitem{Argan:1980zz}
  P.~Argan {\it et al.},
  Phys.\ Rev.\  C {\bf 21} (1980) 1416.

\bibitem{Argan:1987dm}
  P.~Argan {\it et al.},
  Phys.\ Lett.\  {\bf 206B} (1988) 4
  [Erratum-ibid.\  B {\bf 213} (1988) 564].

\bibitem{Bergstrom:1998zz}
  J.~C.~Bergstrom, R.~Igarashi, J.~M.~Vogt, N.~Kolb, R.~E.~Pywell, D.~M.~Skopik and E.~J.~Korkmaz,
  Phys.\ Rev.\  C {\bf 57} (1998) 3203.

\bibitem{Barnett:2008zz}
  M.~G.~Barnett, R.~Igarashi, R.~E.~Pywell and J.~C.~Bergstrom,
  Phys.\ Rev.\  C {\bf 77} (2008) 064601.

\bibitem{Djukanovic:2004px}
  D.~Djukanovic, M.~R.~Schindler, J.~Gegelia and S.~Scherer,
  Phys.\ Rev.\ D {\bf 72} (2005) 045002.
  [hep-ph/0407170].

\bibitem{Kvinikhidze:2009be}
  A.~N.~Kvinikhidze, B.~Blankleider, E.~Epelbaum, C.~Hanhart and M.~P.~Valderrama,
  Phys.\ Rev.\ C {\bf 80} (2009) 044004.
  [arXiv:0904.4128 [nucl-th]].


\bibitem{Epelbaum:2003gr}
  E.~Epelbaum, W.~Gl\"ockle, U.-G.~Mei\ss ner,
  Eur.\ Phys.\ J.\  A {\bf 19 } (2004)  125.
  [nucl-th/0304037].

\bibitem{Epelbaum:2003xx}
  E.~Epelbaum, W.~Gl\"ockle, U.-G.~Mei\ss ner,
  Eur.\ Phys.\ J.\  A {\bf 19 } (2004)  401.
  [nucl-th/0308010].


\bibitem{Liebig:2010ki}
  S.~Liebig, V.~Baru, F.~Ballout, C.~Hanhart and A.~Nogga,
  Eur.\ Phys.\ J.\ A {\bf 47} (2011) 69
  [arXiv:1003.3826 [nucl-th]].

\bibitem{Noggaprivate}
A.~Nogga, private communication.


\bibitem{Vegas}
G.P.~Lepage, J.\ Comp.\ Phys.\ {\bf 27} (1978) 192.

\bibitem{Lepage:1997cs}
  G.~P.~Lepage,
  arXiv:nucl-th/9706029.

\bibitem{Epelbaum:2009sd}
  E.~Epelbaum and J.~Gegelia,
  Eur.\ Phys.\ J.\  A {\bf 41} (2009) 341
  [arXiv:0906.3822 [nucl-th]].

\bibitem{FernandezRamirez:2009jb}
  C.~Fernandez-Ramirez, A.M.~Bernstein and T.W.~Donnelly,
  Phys.\ Rev.\  C {\bf 80} (2009) 065201
  [arXiv:0907.3463 [nucl-th]].

\bibitem{Krebs:2003}
H.~Krebs, ``Neutral pion electroproduction off the deuteron'',
Dissertation, University of Bonn (2003).

\bibitem{Bernard:1996bi}
  V.~Bernard, N.~Kaiser and U.-G.~Mei{\ss}ner,
  Nucl.\ Phys.\ A {\bf 607} (1996) 379
   [Erratum-ibid.\ A {\bf 633} (1998) 695]
  [hep-ph/9601267].

\bibitem{Bernard:1994dt}
  V.~Bernard, N.~Kaiser and U.-G.~Mei{\ss}ner,
  Phys.\ Rev.\ Lett.\  {\bf 74} (1995) 3752
  [hep-ph/9412282].

\bibitem{Bernard:2005dj}
  V.~Bernard, B.~Kubis and U.-G.~Mei\ss ner,
  Eur.\ Phys.\ J.\ A {\bf 25} (2005) 419
  [nucl-th/0506023].


\bibitem{Schmidt:2001vg}
  A.~Schmidt {\it et al.},
  Phys.\ Rev.\ Lett.\  {\bf 87} (2001) 232501
  [arXiv:nucl-ex/0105010].

\bibitem{Gasparyan:2010xz}
  A.~Gasparyan and M.~F.~M.~Lutz,
  Nucl.\ Phys.\  A {\bf 848} (2010) 126
  [arXiv:1003.3426 [hep-ph]].

\bibitem{Rekalo:2001he} 
  M.~P.~Rekalo and E.~Tomasi-Gustafsson,
  Phys.\ Rev.\ C {\bf 66} (2002) 015203
  [nucl-th/0112063].

\bibitem{Baru:2004kw} 
  V.~Baru, C.~Hanhart, A.~E.~Kudryavtsev and U.~G.~Mei{\ss}ner,
  Phys.\ Lett.\ B {\bf 589} (2004) 118
  [nucl-th/0402027].

\end{thebibliography}
\end{document}